\documentclass[12pt]{article}
\usepackage{graphics}
\usepackage{epsfig}
\usepackage{amsmath}

\begin{document}

\setlength{\textheight}{240mm}
\voffset=-15mm
\baselineskip=20pt plus 2pt
\renewcommand{\arraystretch}{1.6}

\begin{center}

{\large \bf  On the energy of Schwarzschild spacetime with the post-Newtonian approximation}\\
\vspace{5mm}
\vspace{5mm}
I-Ching Yang  \footnote{E-mail:icyang@nttu.edu.tw}

Department of Applied Science, National Taitung University, \\
Taitung 95002, Taiwan (R.O.C.)\\

\end{center}
\vspace{5mm}

\begin{center}
{\bf ABSTRACT}
\end{center}
With the post-Newtonian approxination, the energy of Schwarzschild spacetime in the Weinberg prescription 
is obtained. The energy for the first post-Newtonian approximation $E^{(1)} = m$ gives the Newtonian 
treatment of Schwarzschild spacetime. However, for the second post-Newtonian approximation, the erergy 
is shown that $E^{(1)}$ adds extra terms $E^{(2)}$ which consist of the energy stored in the configuration 
$E_{\rm config}$, in the gravitational field $E_{\rm field}$. and a term of surface integral.  These extra terms 
gives post-Newtonian corrections to the Newtonian treatment.

\vspace{2cm}
\noindent
{PACS No.: 04.20.Cv, 04.25.Nx \\}

\vspace{5mm}
\noindent

\newpage

In special relativity, the physical quantity ``mass" has two meanings. The ``invariant mass" $m_0$ (also called rest 
mass) is, in effect, the resistance that a body of matter offers to change in its velocity upon the application of a 
force, and the ``relativistic mass" $m_{\rm ref}$ depends on the observer's inertial frame of reference for an object 
moving at a velocity $\vec{v}$ relative to the observer. Although, general relativity does not offer a single 
definition of the term mass, but offers several different definitions that are applicable under different circumstances,
like ADM~\cite{ADM62} and Bondi masses~\cite{BBM62} in asymptotically flat spacetimes, and Komar mass~\cite{K59} 
in stationary spacetimes, etc. Because of mass-energy equivalence, the concepts of enerrgy and mass are indistinguishable.
So, similar tricky situation also would come about the definition of energy in general relativity.

There are two famous ideas  on the definition of the localized energy-momentum associated with the gravitational field, 
which are energy-mometum complex and quasilocal energy. 
One idea, called {\it energy-momentum complex}, is introduced
\begin{equation}
\Theta^{\mu}_{\nu} = \sqrt{-g} \left( T^{\mu}_{\nu} + t^{\mu}_{\nu} \right)
\end{equation}
which satisfies the differential conservation form $\partial_{\nu} \Theta^{\mu}_{\nu} =0$.  
Here $T^{\mu}_{\nu}$ is the energy-momentum tensor of matter and $t^{\mu}_{\nu}$ is the energy-momentum 
pseudotensor from the gravitational field.
There have been many proposals of energy-momentum complex including Einstein~\cite{E15}, Tolman~\cite{T30}, 
Papapetrou~\cite{P48}, Landau-Lifshitz~\cite{LL62}, M{\o}ller~\cite{M} and Weinberg~\cite{W72} tec.
Another ideaa, {\it quasilocal energy}, associated for a finite region is determined by the integral
\begin{equation}
H( {\bf N}, \Sigma ) = \int_{\Sigma} N^{\mu} {\cal H}_{\mu} +\oint_{\partial \Sigma} {\cal B} ({\bf N}) ,
\end{equation}
which generates the spacetime displacement of a finite spacelike hypersurface $\Sigma$ along a vector field $N^{\mu}$.
Various physicists , such as Hawkin~\cite{H68}, Penrose~\cite{P82}, Brown and York~\cite{BY93}, Hayward~\cite{H94} 
and Nester et al.~\cite{N95} etc.,  had given different definitions for quasilocal energy.
Furthermore, some studies were given the mathermatical relations~\cite{Y04,V06,Y09} and Legendre transformation~\cite{Y12,YCT12,YH14,Y17} 
between Einstein and M{\o}ller energy-momentum complexes in specific spacetime
Im my recent study, according to the first post-Newtonian approximation of Schwarzschild metric, 
the energy components of the Einstein and M{\o}ller energy-momentum complex were obtained,
which involve the rest-mass energy $m$, the energy stored in the configuration and that in the gravitational field.
Thus, in this article, I will estimate the energy component of the Weinberg energy-momentum complex
with the first and second post-Newtonian approxination of Schwarzschild metric.

In oder to suppose a coordinate system that is quasi-Minkowskian, the metric was given as 
\begin{equation}
g_{\mu \nu} = \eta_{\mu \nu} + h_{\mu \nu}  ,
\end{equation}
where $\eta_{\mu \nu}$ is the Minkowski metric. 
Weinberg adopt the energy-momentum pseudotensor of the gravitational field  
\begin{equation}
t_{\mu \nu} = \frac{1}{8\pi} \left[ R_{\mu \nu} - \frac{1}{2} g_{\mu \nu} R^{\lambda}_{\lambda}
 - {R^{(1)}}_{\mu \nu} + \frac{1}{2} \eta_{\mu \nu} {R^{(1)}}^{\lambda}_{\lambda} \right]  ,
\end{equation}
and the part of the Ricci tensor linear in $h_{\mu \nu}$  
\begin{equation}
{R^{(1)}}_{\mu \nu} = \frac{1}{2} \left( \frac{\partial^2 h^{\lambda}_{\lambda}}{\partial x^{\mu} \partial x^{\nu}}
- \frac{\partial^2 h^{\lambda}_{\mu}}{\partial x^{\lambda} \partial x^{\nu}}
- \frac{\partial^2 h^{\lambda}_{\nu}}{\partial x^{\mu} \partial x^{\lambda}} 
+ \frac{\partial^2 h^{\mu}_{\nu}}{\partial x^{\lambda} \partial x^{\lambda}} \right)  .
\end{equation}
Hence the Winberg energy-momentum complex~\cite{W72} can be written as
\begin{equation}
\Theta^{\mu \nu} =  {R^{(1)}}^{\mu \nu} - \frac{1}{2} \eta^{\mu \nu} {R^{(1)}}^{\lambda}_{\lambda} 
= \partial_{\rho} Q^{\rho \mu \nu}
\end{equation}
with Winberg's superpotential is 
\begin{equation}
Q^{\rho \mu \nu} = \frac{1}{2} \left[ \frac{\partial h^{\lambda}_{\lambda}}{\partial x^{\mu}} \eta^{\rho \nu}
- \frac{\partial h^{\lambda}_{\lambda}}{\partial x^{\rho}} \eta^{\mu \nu} - \frac{\partial h^{\lambda}_{\mu}}{\partial x^{\lambda}} \eta^{\rho \nu}
+ \frac{\partial h^{\lambda}_{\rho}}{\partial x^{\mu}} \eta^{\mu \nu} + \frac{\partial h^{\mu}_{\nu}}{\partial x^{\rho}} 
- \frac{\partial h^{\rho}_{\nu}}{\partial x^{\mu}} \right]
\end{equation}
Thus, the energy component of the Weinberg energy-momentum complex within the region $\Sigma$ is shown
\begin{equation}
\begin{split}
E & =  \frac{1}{8\pi} \int_{\Sigma} \Theta^{00} d^3 x = \frac{1}{8\pi} \int_{\Sigma} \frac{\partial}{\partial x^i} Q^{i00} d^3 x  , \\
& =  \frac{1}{16\pi} \int_{\Sigma} \frac{\partial}{\partial x^i} \left( \frac{\partial h_{ik}}{\partial x^k}-
\frac{\partial h_{kk}}{\partial x^i} \right) d^3 x . 
\end{split}
\end{equation}

For a weak-field case, the first post-Newtonian approximation of the Schwarzschild metric~\cite{W72} is studied
\begin{equation}
\begin{split}
g_{00} & =  1 -2\phi + {\cal O}(\phi^2)  \\
g_{0i} & =  0 \\
g_{ij} & =  (-1 - 2\phi) \delta_{ij} + {\cal O}(\phi^2)  ,
\end{split}
\end{equation}
and the qualities $h_{\mu \nu}$ are given as 
\begin{equation}
\begin{split}
h_{00} & = -2\phi + {\cal O}(\phi^2)  \\
h_{ij} & =  -2\phi \delta_{ij} + {\cal O}(\phi^2)  ,
\end{split}
\end{equation}
where $\phi$ is the Newton's gravitational potential and normalized such that $\phi(\infty)=0$.
Then Winberg's superpotential is obtained
\begin{equation}
Q^{i00} = 2 \nabla \phi  ,
\end{equation}
and the energy component of the Weinberg energy-momentum complex is described to
\begin{equation}
{}_{W} \Theta^0_0 = \frac{1}{4\pi} \nabla^2 \phi .
\end{equation}
Therefore, in the Winberg prescription, the energy within the region $\Sigma$ that includes mass distribution is given by
\begin{equation}
E = \frac{1}{4\pi} \int_{\Sigma} \nabla^2 \phi d^3 x  .
\end{equation}
Since the gravitational potential satisfes the field equation with a matter density $\rho$
\begin{equation}
\nabla^2 \phi = 4\pi \rho
\end{equation}
and the total mass is given by 
\begin{equation}
\int_{\Sigma} \rho d^3 x = m .
\end{equation}
Such, in first post-Newtonian approximation,  the energy in the Winberg prescription can be rewritten
\begin{equation}
E = m  ,
\end{equation}
and the energy is contributed by the ``rest-mass energy" as the invariant mass is the rest energy in special relativity.

To study the contribution other than the rest-mass energy, the second post-Newtonian approximation of the Schwarzschild 
metric~\cite{W72} is considered
\begin{equation}
\begin{split}
g_{00} & = 1 -2\phi + 2\phi^2 +{\cal O}(\phi^3)  , \\
g_{0i} & = 0 , \\
g_{ij} & = (-1 - 2\phi -\phi^2) \delta_{ij} - \phi^2 \frac{x_i x_j}{r^2} +
{\cal O}(\phi^3)  ,
\end{split}
\end{equation}
and the qualities $h_{\mu \nu}$ are obtained as 
\begin{equation}
\begin{split}
h_{00} & = -2\phi + 2\phi^2 +{\cal O}(\phi^3)  , \\
h_{ij} & = (-2\phi -\phi^2) \delta_{ij} - \phi^2 \frac{x_i x_j}{r^2} +
{\cal O}(\phi^3)  .
\end{split}  .
\end{equation}
Here the Winberg's superpotential is computed 
\begin{equation}
Q^{i00} = 2 \nabla \phi + 2 \phi \nabla \phi - \frac{2\phi^2}{r} \hat{r} ..
\end{equation}
The energy component of the Møller energy-momentum complex is obtained
\begin{equation}
{}_{W} \Theta^0_0 = \frac{1}{4\pi} \nabla^2 \phi + \frac{1}{4\pi} \nabla \cdot (\phi \nabla \phi)
- \frac{1}{4\pi} \nabla \cdot \frac{\phi^2}{r} \hat{r} 
\end{equation}
and the energy within the region $\Sigma$ is followed 
\begin{equation}
E= \frac{1}{4\pi} \int_{\Sigma} \nabla^2 \phi d^3x - 
\frac{1}{4\pi} \int_{\Sigma} \nabla \cdot ( \phi \nabla \phi ) d^3 x  .
+\frac{1}{8\pi} \int_{\Sigma} \nabla \cdot \left( \frac{\phi^2}{r} \right) d^3x  .
\end{equation}
Although, the second term of the right-hand side in Eq. (19) can be rewriiten as
\begin{equation}
\frac{1}{4\pi} \int_{\Sigma} \nabla \cdot ( \phi \nabla \phi) d^3 x
= \frac{1}{4\pi} \int_{\Sigma} \phi \nabla^2 \phi d^3 x + \frac{1}{4\pi} \int_{\Sigma} (\nabla \phi)^2 d^3 x  .
\end{equation}
and the first term of right-hand side in Eq.(20) is the energy stored in the configuration
\begin{equation}
E_{\rm config} \equiv \frac{1}{8\pi} \int_{\Sigma} \phi \nabla^2 \phi d^3  ,
\end{equation}
and the second term is the energy stored in the gravitational field 
\begin{equation}
E_{\rm field} \equiv \frac{1}{8\pi} \int_{\Sigma} (\nabla \phi)^2 d^3 x  .
\end{equation}
The third term of the right-hand side in Eq.(19) can be expressed in terms of surface integral 
\begin{equation}
\frac{1}{8\pi} \int_{\Sigma} \nabla \cdot \left( \frac{\phi^2}{r} \right) d^3x =
\frac{1}{8\pi} \oint_{\partial \Sigma} \frac{\phi^2}{r} \hat{r} \cdot d\hat{a}  .
\end{equation}
Consequently, in second post-Newtonian approximation, the energy in the Winberg prescription within the
region $\Sigma$ is shown as
\begin{equation}
E = m + 2E_{\rm config} + 2E_{\rm field} - \frac{1}{8\pi} \oint_{\partial \Sigma} \frac{\phi^2}{r} \hat{r} \cdot d\hat{a}  ,
\end{equation}
and resembles the energy in the Einstein prescription
\begin{equation}
E_{\rm E} = m + E_{\rm config} + E_{\rm field} + \frac{1}{8\pi} \oint_{\partial \Sigma} \frac{\phi^2}{r} \hat{r} \cdot d\hat{a}  .
\end{equation}
                                  
In summary, the energy of the Schwarzschild spacetime for the first post-Newtionian approximation by using the 
Weinberg energy-momentum complex is obtained
\begin{equation}
E^{(1)} = m ,
\end{equation}
and gives the Newtonian treatmenet of Schwarzschild spacetime~\cite{MTW}. The energy for the second post-Newtionian 
approximation is that $E^{(1)}$ adds extra terms 
\begin{equation}
E^{(2)} = 2 E_{\rm config} + 2E_{\rm field} - \frac{1}{8\pi} \oint_{\partial \Sigma} \frac{\phi^2}{r} \hat{r} \cdot d\hat{a} ,
\end{equation}
and $E^{(2)}$ gives post-Newtonian corrections to the Newtonian treatemnt~\cite{MTW}. 
Referrung to quantum electrodynamics, $E^{(1)}$  and $E^{(1)} + E^{(2)}$ behave like bare mass and experimentally observable 
masss, and $E^{(2)}$ is the increase in mass owing to the interaction of the matter with the field. 
Particularly, the term of surface integral in Eq.(29) could be 
suggested as the flux of $(\phi^2 /r) \hat{r}$ through a surface $\partial \Sigma$. Thus, for energy-momentum complexes, to 
study the energy in higher-order post-Newtonian approximation will be useful to clarify their physical meanings.

\end{document}